\documentclass[%
 reprint,
 amsmath,amssymb,
 aip,jcp
]{revtex4-2}

\usepackage{graphicx}% Include figure files
\usepackage{dcolumn}% Align table columns on decimal point
\usepackage{bm}% bold math
\usepackage[table]{xcolor}
\usepackage{graphicx}
\usepackage{adjustbox}
\usepackage{placeins}
\usepackage[T1]{fontenc}
\usepackage{lipsum}
\usepackage{csquotes}
\usepackage{colortbl}
\usepackage{hyperref}
\usepackage{bm}
\usepackage{multirow}
\usepackage{booktabs}

\newcommand{\comment}[1]{\textcolor{black}{#1}}

\begin{document}

\title{Perspective on integrating machine learning into computational chemistry and materials science}

\author{Julia Westermayr}
\affiliation{Department of Chemistry, University of Warwick, Gibbet Hill Road, Coventry, CV4 7AL, United Kingdom}
\author{Michael Gastegger}
\affiliation{Machine Learning Group, Technische Universit\"at Berlin, 10587 Berlin, Germany}
\author{Kristof T. Sch\"utt}
\affiliation{Machine Learning Group, Technische Universit\"at Berlin, 10587 Berlin, Germany}
\affiliation{Berlin Institute for the Foundations of Learning and Data, 10587 Berlin, Germany}
\author{Reinhard J. Maurer}
\email{r.maurer@warwick.ac.uk}
\affiliation{Department of Chemistry, University of Warwick, Gibbet Hill Road, Coventry, CV4 7AL, United Kingdom}


\date{\today}% It is always \today, today,

\begin{abstract}
Machine learning (ML) methods are being used in almost every conceivable area of electronic structure theory and molecular simulation. In particular, ML has become firmly established in the construction of high-dimensional interatomic potentials. Not a day goes by without another proof of principle being published on how ML methods can represent and predict quantum mechanical properties -- be they observable, such as molecular polarizabilities, or not, such as atomic charges. As ML is becoming pervasive in electronic structure theory and molecular simulation, we provide an overview of how atomistic computational {modeling} {is} being transformed by the incorporation of ML approaches. 
%old
From the perspective of the practitioner in the field, we assess how common workflows to predict structure, dynamics, and spectroscopy are affected by ML. Finally, we discuss how a {tighter} and lasting integration of ML methods with computational chemistry and materials science can be achieved and what it will mean for research practice, software development, and postgraduate training.
\end{abstract}

\maketitle

%\tableofcontents

%%%FOCUS on combining ML with QM/EST to make practically useful advances
\section{\label{sec:introduction}Introduction}
\begin{figure}[tb]
    \centering
    \includegraphics[width=.95\columnwidth]{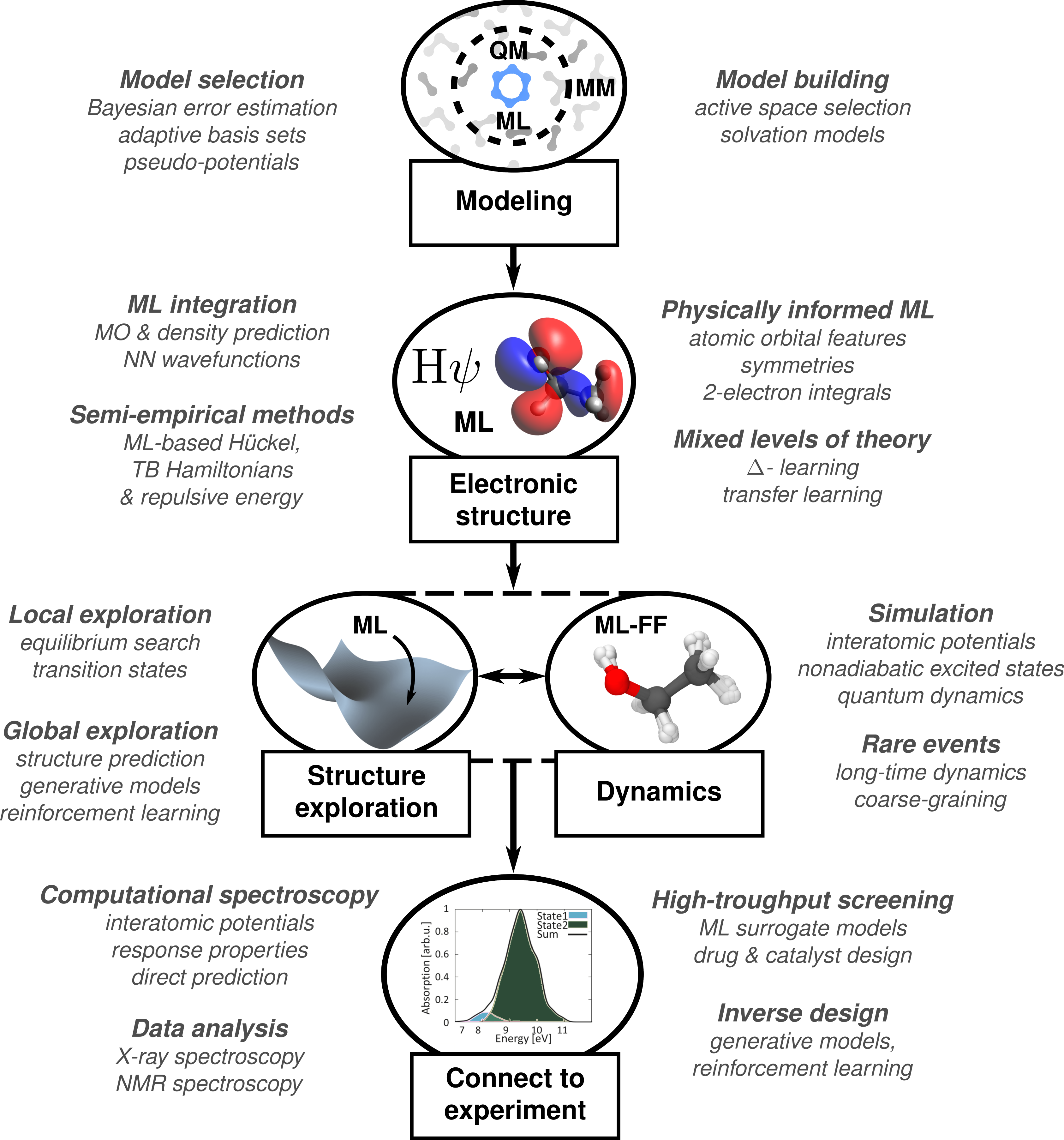}
    \caption{\label{fig:1}Schematic depiction of the key workflow steps in computational molecular and materials {modeling}: Model building and method choice, electronic structure calculations, structure exploration and dynamics, and connection to experiment. All of these steps can benefit from ML models. In many cases ML methods do not just enhance existing approaches, but also open avenues {toward} new workflows.
    }
\end{figure}

%PAR 1 The status quo
Atomistic and electronic structure simulations based on quantum theoretical calculations form a central aspect of modern chemistry and materials research. {They enable the prediction of molecular and materials properties from first-principles as well as the simulation of atomic-scale dynamics. On this basis, computational chemists and physicists in academia and industry contribute to fundamental mechanistic understanding of chemical processes, to the identification of novel materials, and the optimization of existing ones.}  Over the last few decades, computational molecular simulation has been firmly established in the chemical sciences as an important part of the method portfolio. This was accompanied by a move to streamline and optimize common workflows for model building and simulation (see Figure \ref{fig:1}). {Algorithms for molecular geometry optimization, efficient molecular dynamics simulations, and electronic structure calculations perform highly specialized tasks while being massively scalable and parallelized across a diverse range of hardware architectures.~\cite{ESL,ELSI}} Simultaneously, PhD graduates in the field have been trained to be expert users of existing and developers of new simulation workflows. This is the \emph{status quo} at the time when machine learning (ML) methods enter the stage.

%PAR 2 
\comment{The application of ML to atomistic simulation and electronic structure theory has been developing rapidly since its earliest works in a modern context.~\cite{behler2007,Carbogno2008PRL,Dawes2008JCP,Manzhos2009,Bartok2010,rupp2012fast,Netzloff2006JCP,Evenhuis2011JCP,Braams2009IRPC}} A number of excellent reviews have recently been written to highlight progress in various contexts including the role of ML in catalyst design,~\cite{freeze2019search,Elton2019} in the development of force-fields and interatomic potentials for ground state properties{~\cite{Behler2017, Mueller2020, Manzhos2020_review, Gkeka2020,Unke2020arXiv,Deringer2019AM}} and excited states,~\comment{\cite{Westermayr2020CR,Westermayr2020MLST_Perspective,Dral2021NRC}} in quantum chemistry,~\cite{Dral2020,von2020exploring} in finding solutions to the Schr\"odinger equation,~\cite{Manzhos2020} and the role of unsupervised learning in atomistic simulation~\cite{Ceriotti2019} (see Table I for a non-exhaustive list). 

%PAR3
An excellent retrospective of the last decade of ML in the context of chemical discovery has recently been published by von Lilienfeld and Burke,~\citep{Lilienfeld2020NC} predicting a bright future in the context of ML for quantum chemistry that lies ahead. 
Indeed, not a day goes by without another novel ML approach being published, which promises to predict atomic and electronic properties of molecules and materials at ever greater accuracy and efficiency. {A main goal of many ML models is the parametrization of analytical models to represent electronic structure. These ML models  can then be evaluated extremely fast. Thus ML models can speed up simulations to achieve longer time and length scales.  Their efficiency depends strongly on the design of descriptors or neural network architectures that optimally chart the vast space of chemical compounds and materials.}~\cite{Lilienfeld2018review,schutt2020machine} These approaches have the potential to fundamentally change day-to-day practices, workflows and paradigms in atomistic and quantum simulation {as they become more tightly integrated with existing tools}. \textbf{But how exactly will ML affect the method portfolio of future computational scientists working in electronic structure theory and molecular simulation?} How will this affect a practitioner who wants to determine the equilibrium structure and ground-state energy of a molecular system using electronic structure theory? How will it change the required expertise and demands on PhD graduates?

%PAR4 Purpose of this review
\comment{For the uninitiated, it is easy to get lost in the vast array of ML models, which might soon be comparable to the zoo of exchange-correlation functionals available in density functional theory (DFT).~\cite{Becke2014JCP}} What will become the ML equivalent of go-to DFT functionals for practitioners? At the moment, there are relatively few examples where ML models have become generally applicable to researchers outside the immediate circle of developers. In this perspective, we are discussing recent advances through the lens of their potential benefit to a wide community of computational molecular scientists {who are not ML experts. Our goal is} to identify future possibilities of permanent integration of ML-based approaches into workflows and electronic structure and simulation software packages. {This can for example involve a common code base and data structure for ML and simulation algorithms or bidirectional data exchange between workflows based on ML or physical simulation.} Central to this perspective is the question how ML can effectively address the computational bottlenecks {and capability gaps} in electronic structure calculations and molecular simulations and what are the steps needed to make ML an integral part of the method portfolio of this field. 

\comment{Our goal is to make this account as accessible as possible and to highlight applications and approaches that the community might want to keep track of in the future.} We stress that our aim is not to provide a comprehensive review of existing {ML descriptors, representations, and} approaches, which is beyond the scope of this perspective {and well covered by further reading} material in Table \ref{tab:my_label}. Following the key steps of molecular {modeling} shown in Figure \ref{fig:1}, each section focuses on how ML methods can benefit a central workflow or aspect of computational molecular and material science {(\textit{cf.} highlighted sentences in each paragraph)}. We place a particular focus on approaches that have the potential to augment existing or introduce new prevalent approaches.

%%%%TABLE
%List a selection of the most recent reviews between 2018 and 2020 in a table
\begin{table}[t]
    \centering
    \begin{tabular}{lll} \toprule 
    \textbf{Year} & \textbf{References} & \textbf{Topic of ML Review} \\ \midrule
     2017 &\citet{Behler2017} & Interatomic Potentials \\
     2018 & \citet{Goldsmith2018} & ML in Catalysis \\     
     2019 & \citet{Carleo2019} & ML in Physical Sciences \\ 
     2019 & \citet{Yang2019CR} & Drug Discovery  \\
     2019 & \citet{Elton2019} & Molecular Design  \\     
     2019 & \citet{Schleder2019} & ML in Materials Science  \\
     2019 & \citet{Ceriotti2019} & Unsupervised Learning \\      
     2020 & \citet{Dral2020} &  ML in Quantum Chemistry   \\
     2020 & \citet{noe2020machine} & Molecular Simulation \\
     2020 & von Lilienfeld \textit{et al.} \cite{von2020exploring} & Chemical Space \\
     2020 & Mueller \textit{et al.} \cite{Mueller2020} & Interatomic Potentials \\
     2020 & Manzhos \textit{et al.} \cite{Manzhos2020_review} & Small Molecules and  Reactions \\
     2020 & \citet{Gkeka2020} & Force Fields \& Coarse Graining \\
     2020 & \citet{Unke2020arXiv} & Force Fields \\ 
     2020 & \citet{Toyao2020_ACS_Catalysis} & Catalysis Informatics  \\ 
     2020 & \citet{Manzhos2020} & ML in Electronic Structure \\ 
     2020 & Westermayr \textit{et al.}\cite{Westermayr2020CR} & ML for Excited States  \\
     2021 & \citet{Behler2021} & Neural Network Potentials  \\     \bottomrule
    \end{tabular}
    \caption{Overview of recent reviews of machine learning methods in electronic structure theory and atomistic simulation. This is not intended to be a complete list of all reviews on the subject, but a selection of suggested further reading.}
    \label{tab:my_label}
\end{table}

%%%%%%%%%%%%%%%%%%%%%%%%%%%%%%%%%%%%%%%%%%%%%%
\section{Machine Learning Primer}\label{sec:mlprimer}

{We start by introducing basic terminology and concepts of ML that will be used in the remaining sections of the perspective.}
ML is concerned with algorithms that improve with increasing amount of available data under some performance measure. 
{Statistical learning theory offers a general framework to find predictive functions $f: \mathcal{X} \rightarrow \mathcal{Y}$ mapping an input space $\mathcal{X}$ to a target space $\mathcal{Y}$.~\cite{hastie2009elements} }
In contrast to conventional physical models, where one often starts with clear assumptions about the system to be modelled, ML focuses on \emph{universal approximators}.
These are able to represent any function with arbitrary accuracy, when given enough training data and parameters.
{Examples for this class of models are Gaussian processes (GPs) or neural networks (NNs).~\cite{leshno1993multilayer}
GPs are defined by linear combinations of the covariances between data points. These are given by a suitable (nonlinear) kernel function.
NNs consist of a sequence of multiple linear transforms, alternated with nonlinear \emph{activation functions}.
This is also referred to as \emph{deep learning}, where each set of transform and nonlinearity is called a \emph{layer}.}

\begin{figure}
    \centering
    \includegraphics[width=0.90\linewidth]{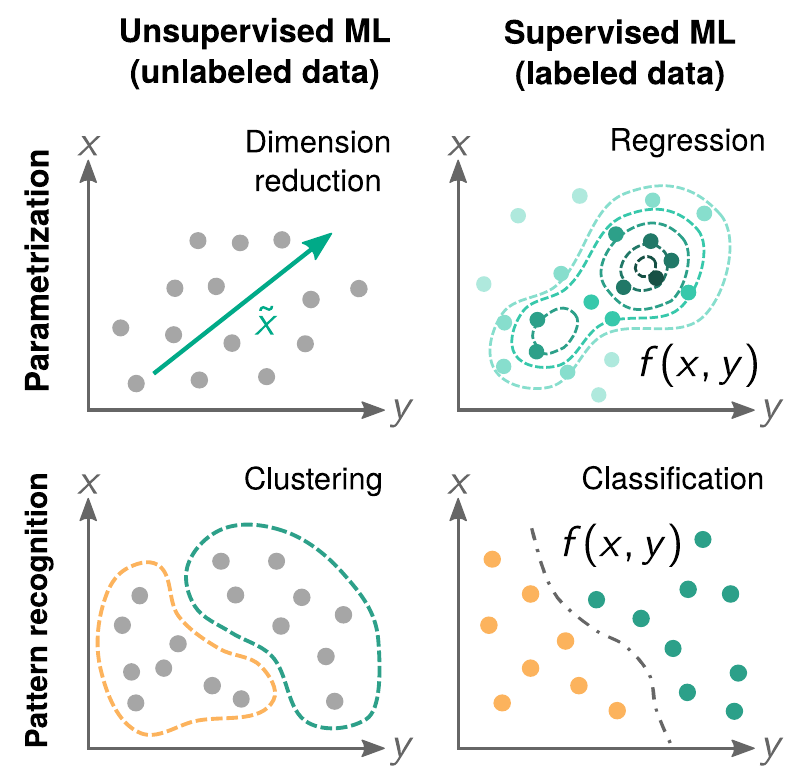}
    \caption{Schematic depiction of different ML model categories. Unsupervised learning techniques use unlabeled data and are often  used for dimensionality reduction or clustering, whereas supervised ML models perform regression or classification tasks on labelled data.}% First draft of schematic depiction of different machine learning approaches categorized by the type of data they work on and the task they perform.}
    \label{fig:2}
\end{figure}

{The functional relationship to be found is specified by choosing a suitable \emph{loss function}.
If the loss $\ell(f(x), y)$ requires knowledge of the targets $y \in \mathcal{Y}$, this is called \textbf{supervised learning}. 
This includes classification and regression for categorical and continuous target spaces $\mathcal{Y}$, respectively (see also Fig. \ref{fig:2}).}
ML force fields are examples of regression tasks (see section \ref{sec:QM}),~\cite{hansen2013assessment} {where often the squared error is used as loss function. }
For instance, classifiers can be used to automatically select appropriate quantum chemistry methods for a given system (see section \ref{sec:model}).
In contrast, \textbf{unsupervised ML} aims to find patterns in the data that are specified by a loss function without having access to the ground truth targets $y$.
{Tasks falling under this category include clustering, dimensionality reduction, or density estimation of the data distribution}.
In the context of computational chemistry, unsupervised ML finds application in post-processing and analysis of {molecular} simulation data, {\textit{e.g.}}{,} in {identifying collective variables (CVs)}and reaction pathways that will be discussed in section \ref{sec:QD} (see also Fig. \ref{fig:2}). 

{The optimal predictive function minimizes the \emph{expected risk}, {\textit{i.e.}}, the expectation of the loss function weighted by the probability distribution over the data.~\cite{muller2001introduction}
However, the data distribution is usually unknown and, in supervised learning, the loss requires access to the targets.
Thus, one instead optimizes the \emph{empirical risk}, {\textit{i.e.}}, the expectation over a training set sampled from the data distribution.
This could for example consist of electronic structure calculations of systems $x \in \mathcal{X}$ with properties $y \in \mathcal{Y}$.
Since there typically exist many possible approximates that fit a finite training set, one introduces regularizer terms to the optimization problem, which punish complex solutions.
This avoids \emph{overfitting}, {\textit{i.e.}}, an increased error on unseen data due to approximating a simple functional relationship with an overly complex function on the training set.}

{Another important aspect to consider is the selection of training examples, which should be representative of the distribution encountered when applying the ML model.
This requires not only a sufficient number of training examples, but also sufficient coverage of the input space.
If an ML model is applied outside of its training domain, {\textit{i.e.}}, if it is used for extrapolation, its predictions quickly become  unreliable.
\emph{Active learning} aims to detect this and acquire additional training data in the corresponding regions.
Similarly, \emph{Bayesian optimization} is an approach for global search that obtains additional examples where there is a high probability to optimize a given criterion based on the current model and its uncertainty.}
\comment{
ML models are typically evaluated on a separate test set that is not used during the training process, \text{i.e.}, also not for controlling overfitting. To get a better measure of the reliability of ML models in different regions and to detect holes, additional sampling of data can be carried out with \textit{e.g.} enhanced sampling techniques.~\cite{Tao2019TCA,Yang2019JCP}
Alternatively, when using two NNs, minima of their negative squared difference surface can be used to detect sparse conformational regions.~\cite{Lin2020JCP}}

{To design accurate and data-efficient ML models, it is important to be aware of the structure of the input space and how it is represented.
Encoding prior knowledge in the model reduces the effective space to cover and, thus, the required amount of training data.
Examples include the use of convolutions to encode roto-translational invariances~\cite{Schutt2017_double} or delta learning, where only the difference to a baseline is learned.~\cite{ramakrishnan2015big}.
Beyond that, \emph{transfer learning} studies how knowledge contained in models trained on one task can be reused for related tasks. This also means that, the question of whether a prediction is an extrapolation depends not only on the given training data, but also on the prior knowledge built into the ML model.}

{By employing a probabilistic input space and a structured target space, one obtains a model that can{,} {\textit{e.g.}}{,} be used to generate novel molecular structures.}
The probability distribution over molecular space can be modeled explicitly, for example using variational autoencoders,~\cite{kingma2013auto} or implicitly{,} {\textit{e.g.}}{,} by generative adversarial networks~\cite{goodfellow2014generative} that provide access to the distribution only through sampling.
In a supervised setting, generative models can facilitate inverse design by learning a probability distribution of chemical structures conditioned on a desired target range of one or multiple properties.

Finally, \textbf{reinforcement learning} is concerned with learning the optimal action in a given state to maximize a specified, future reward.
An example for this is an unfolded protein (state), where one applies changes to the geometry (action) in order to come closer to the folded structure with minimum energy (future reward).~\cite{shamsi2018reinforcement}
Reinforcement learning includes an exploration strategy such that more data is collected during the training process.
Therefore, it can, for example, be used for molecular design without requiring a representative set of reference structures before training.

%%%%%%%%%%%%%%%%%%%%%%%%%%%%%%%%%%%%%%%%%%%%%%
\section{ML improves model building, method choice, and opens new  multi-scale approaches}\label{sec:model}

The first task one faces when investigating a chemical problem \emph{in silico} is to determine a suitable computational model. The modeling process involves the design of the atomistic structural model and the choice of computational method for calculating the properties of interest. 
Both choices traditionally are based on achieving a balance between a sufficiently accurate description of the chemical phenomena to be studied and limited computational effort that renders the calculations feasible.

{Computational methods can range from electronic structure theory methods ({\textit{e.g.}}{,} correlated wavefunction or density functional approaches) to more approximate empirical force fields. Depending on the level of approximation, a method can be appropriate for modeling certain phenomena, while being less reliable for others. One example are classical empirical force fields, which sacrifice the ability to model chemical {bond breaking in favor of computational speed, but yield excellent predictions for ensemble averages of macromolecular systems. 
Different applications also place different accuracy requirements on the reference method.
A concept often mentioned in the context of ML in chemistry is chemical accuracy, which originally  specified that the energy error of a computational method deviates at most 1 kcal/mol from experiment. 
This accuracy requirement was coined by Pople in his Noble lecture~\cite{PopleNobel} for thermodynamic properties, where it allows reliable comparison with experiment.~\cite{Curtiss2000JCP}
However, other applications may necessitate significantly more rigorous error limits.
In the field  of high-resolution vibrational spectroscopy for example, reliable predictions require so-called spectroscopic accuracy, which corresponds to an energy error smaller than 1 cm$^{-1}$ or 0.003 kcal/mol.~\cite{Puzzarini2019CR}}
The model building stage furthermore involves a range of decisions on how to represent the system, for example, how to treat environments such as solvents, what  size the simulation cell should have, or which atoms to model explicitly. All these decisions can influence the quality of results at a fundamental level and hence need to be considered carefully.}

{Unfortunately, choices are often ambiguous and different strategies can still yield similar results or may only work in certain combinations. The associated design choices typically require a mix of expertise and chemical intuition of experienced practitioners. This makes it hard to see how ML could help to automate this process. Nevertheless, ML models can{,} {\textit{e.g.}}{,} learn to infer decision rules or categorize complex patterns in a purely data driven fashion. This makes them a promising tool to provide support during the model building stage, making balanced model building choices more widely available and potentially achieving fully automated decision making in the future.}

% uncertainty
{Transparent method selection protocols can be based on \textbf{uncertainty quantification}.}~\cite{ruscic2014uncertainty,chernatynskiy2013uncertainty}
Currently, theoretical predictions tend to be reduced to a single number, without considering the spread due to{,} {\textit{e.g.}}{,} method-specific modeling errors.
Access to confidence intervals can provide several key advantages beyond determining how well a particular method is suited for a task.
Trends in method predictions can be analyzed in a more general manner, going beyond the snapshots provided by traditional benchmark studies.
When combined with experiment, uncertainties assigned to theoretical predictions allow for a better separation of error sources and interpretation of results. 
Recently, some progress has been made in tackling this problem with ML algorithms and Bayesian approaches in particular.  
% DFT
Bayesian error estimation has been successfully used to construct multiple density functionals.
\citet{wellendorff2012density} reported a Bayesian functional with a non-local van der Waals {(vdW)} correlation term. This so-called BEEF-vdW functional provides predictions as well as computational error estimates.
They demonstrated the utility of BEEF-vdW based on two surface science problems, modeling graphene adsorption on a Ni(111) surface and the binding of CO to Pt(111) and Rh(111) substrates.
Bayesian frameworks for density functionals were also developed by \citet{aldegunde2016development} and \citet{simm2016systematic}.
All these approaches allow for the construction of specialized density functionals which yield confidence intervals for computed energies. 
This makes it possible to automatically probe the reliability of the method for different compounds and structures and identify problematic situations.
{\citet{simm2016systematic} used their approach to estimate the errors associated with different reaction barriers along the catalytic cycle of Yandulov--Schrock catalyst, where they demonstrated that even similar reaction steps can exhibit very different confidence levels due to shortcomings of the computational method.}
By applying this approach to chemical reaction networks, \citet{proppe2017uncertainty} demonstrated how this method can further be used to provide uncertainty estimates for chemical reaction rates.

% Basis Set
Beyond error estimates, ML has been employed to automatically construct basis sets for electronic structure methods.~\cite{schuett2018machine} 
{Usually, pre-defined basis sets are used for electronic structure computations, which aim to provide reasonable accuracy over a wide range of compounds. As such they use higher radial and angular resolution than might be necessary for certain molecules.
\citet{schuett2018machine} have shown how ML can be used to generate an adaptive basis set tailored  to a specific system based only on local structural information.
Using liquid water as example, their adaptive basis set was able to reduce computational cost by up to a factor of 200.}
% Adaptive basis sets on the other hand are tailored to a specific system based only on local structural information and have been shown to significantly improve the accuracy and efficiency of subsequent electronic structure calculations. 
Similarly, local pseudopotentials have been constructed based on kernel ridge regression.~\cite{Lueder2020}
% MR character 
Another important decision in method selection is whether the problem of interest exhibits strong electron correlation (also referred to as multi-reference character or static correlation). In this case, a single {antisymmetric product wave function} is no longer sufficient to describe the {electronic} system and single-reference methods ({\textit{e.g.}}{, semi-local approximations to DFT, single-reference {coupled-cluster (CC)} theory) yield inconsistent performance across configurational space and fail to describe bond breaking.} \citet{Duan2020JPCL} have proposed a {semi-supervised} ML approach to automatically classify chemical systems according to their multi-reference character in an efficient manner. {This makes it possible to identify problematic systems without the need to carry out expensive high-level calculations and thus aid in the method selection process.}
% Composite methods
In some situations, it can be advantageous to rely not on a single method, but instead employ a combination of electronic structure theories and basis set levels. Such composite methods {have a long history in  computational chemistry, with the Gaussian methods for thermochemistry (G2-G4)~\cite{Curtiss1991JCP,Curtiss1998JCP,Curtiss2007JCP} being some of the most prominent examples. All composite methods have in common, that they} profit from the cancellation of errors at different levels of theory and can offer improved accuracy at lower computational cost. \citet{zaspel2018boosting} have leveraged ML and combination techniques to derive a composite method in a data driven fashion. They could demonstrate that their method achieved CC accuracy using only lower levels of theory.
 
The \textbf{model building} process encompasses many other aspects apart from method selection. This includes decisions on which structural aspects of the system need to be {considered explicitly or only accounted for in their implicit effect on the system} ({\textit{e.g.}}{,} implicit versus explicit solvation models), whether periodic boundary conditions are required or which boundary box shapes and sizes are appropriate. Other aspects concern the electronic structure, especially in the context of multi-reference methods. Most of these approaches require decisions on which particular electronic reference configurations, often referred to as active space, to include in the description of a system. This problem is highly nontrivial, as it not only depends on the intrinsic electronic structure of a system but also on the chemical {reaction} to be studied. As a consequence, these methods ({\textit{e.g.}}{, Complete Active Space Self Consistent Field (CASSCF))} have been hard to use {by non-expert users} in a black box manner {in the past}. 
% active space selection
\citet{Jeong2020JCTC} recently introduced a ML protocol {based on decision trees} for active space selection in bond dissociation studies. Their approach is able to predict active spaces {able to reproduce the dissociation curves of diatomic molecules} with {a success rate of approximately 80 percent precision compared to random selection. This} constitutes an important step {toward} black box applications of multi-reference methods.

% multi scale
ML approaches further show great potential in the context of \textbf{multi-scale modeling}.
Multi-scale approaches combine information from different levels of theory to bridge different physical scales.
Examples include hybrid quantum mechanics/molecular mechanics (QM/MM) simulations~\cite{zhang2018potential}. For example, \citet{zhang2018solvation} have shown how a simple $\Delta$-learning based model can improve the accuracy of solvent free energy calculations{, where they could reach hybrid DFT accuracy using a semi-empirical DFTB baseline.}
A similar scheme has been employed by \citet{boselt2020machine} to simulate the interactions of organic compounds in water.
\citet{gastegger2020machine} used a ML/MM approach where a ML model completely replaced the QM region to model solvent effects on molecular spectra and reactions.
{This made it possible to achieve an acceleration of up to four orders of magnitude, while still retaining the accuracy of the hybrid functional reference method.}
Combining fragment methods with ML techniques, \citet{Chen2019JPCL} were able to investigate excited states in extended systems {in an efficient manner by only treating the photochemically active region with a multi-reference method while the environment is modeled with ML.}
Finally, \citet{caccin2015framework} have introduced a general framework for leveraging multi-scale models using ML to simulate crack propagation through materials{, thus enabling simulations which would otherwise be impossible using either classical force-fields or electronic structure methods alone.}

{\textit{Future directions:} 
% final statement and perspective
While a complete automation of the model building stage has not yet been achieved, ML based algorithms have nevertheless led to significant progress toward this endeavor. Due to the complexity of the model building process, there still is a large number of untouched subjects which may serve as fruitful substrate for future ML research.
Potential avenues include the automated selection of suitable levels of correlation methods for specific problems and using ML to automatically generate partitions in multi-scale approaches.
}

%%%%%%%%%%%%%%%%%%%%%%%%%%%%%%%%%%%%%%%%%%%%%%%%%%%%%%%%%%%%%%%
\section{\label{sec:QM}ML in electronic structure theory}

The solution to the electronic Schr\"odinger equation can be approximated in various ways, where a tug-of-war between accuracy and computational efficiency is crucial to any choice of method. 
The bottlenecks that need to be addressed to achieve more efficient electronic structure calculations are mainly:
\begin{itemize}
    \item[(1)] the evaluation of multi-centre and multi-electron interaction integrals, which requires optimally-tuned basis representations to construct Hamiltonians and sets of secular equations and 
\item[(2)] the (iterative) solution of coupled sets of equations to predict total energies, wave functions, electron densities, and other properties derived thereof.
\end{itemize}
To overcome these bottlenecks, developments of correlated wave-function-based methods, exchange-correlation functionals within DFT, and methods based on many-body perturbation theory must go hand in hand with algorithmic advances. Progress on challenge (2) has been propelled by algorithmic ingenuity and a collective community effort to develop massively scalable linear algebra algorithms to be collected in central libraries such as the Electronic Structure Library (\hyperlink{https://esl.cecam.org/}{ESL}~\cite{ESL}) and the Electronic Structure Interface (\hyperlink{https://wordpress.elsi-interchange.org/}{ELSI}~\cite{ELSI}). It is challenge (1), where ML methods can potentially have the biggest impact in eliminating computational bottlenecks while maintaining high predictive power. 

% TASK 1 very briefly start with interatomic potentials
Currently, the most pervasive application of ML is \textbf{to replace \emph{ab-initio} electronic structure calculations with \emph{ab-initio}-quality interatomic potentials}. {In doing so, ML methods also significantly improve the predictive capabilities of molecular dynamics {(MD)} simulations by enabling \emph{ab-initio}-accuracy at computational costs comparable to classical force fields (\textit{cf.} section~\ref{sec:QD}).}
In principle, ML models can parametrize any {smooth} function, such as the ground-state total energy, the forces, and other derived properties obtained from a first-principles calculation. 
%This eliminates the previously described computational bottlenecks and key targets that are often parametrized with ML models are potential energies, forces, and dipole moments.
Related ML models for {interatomic potentials have already been} reviewed extensively (see Table \ref{tab:my_label} for example). {We therefore focus on ML representations of electronic structure quantities beyond ground-state energies and forces in the following.}

%TASK 2 excited state stuff
\textbf{Many ML representations of excited state properties}, such as HOMO-LUMO gaps,~\cite{Pronobis2018EPJB,Ghosh2019AS,schutt2019unifying} excited-state energies,{~\cite{Ramakrishnan2015,Westermayr2020MLST_Perspective,Westermayr2020JCP,Westermayr2021physically}} or band gaps \cite{schutt2014represent,zhuo2018predicting,Lee2016PRB,Pilania2017CMS} have been proposed {and were mainly based on NNs or kernel methods.}
Recently, ML models have also been applied to derive {excited-state or response properties explicitly by learning the density of states~\cite{Mahmoud2020PRB} or orbital energies,~\cite{Ghosh2019AS,Westermayr2021physically} respectively. These models have further been applied to obtain excitation spectra.} 
{However, a main} challenge that is frequently encountered when fitting {many energy levels} is 
the non-smoothness of the target functions, {which is true for orbital energies as well as adiabatic potential energy surfaces (PESs)}.~\cite{Westermayr2019CS,Duan2020JPCL} Avoided crossings {at conical intersections in adiabatic potential energy landscapes} represent a good example for this behaviour: When two potential energies become degenerate and form a cusp, the respective coupling values become singular at this point in the conformational space. Consequently, a direct learning of such properties is prohibited {in many cases, making a smoothing of the target property or novel fitting approaches preferable. Approaches to achieve better learning behaviour strongly depend on the purpose of the ML model. For instance, in case of spectroscopic predictions it is sufficient to learn the spectral shape directly instead of the energy levels. This has been done with Gaussian Approximation Potentials for the density-of-states~\cite{Mahmoud2020PRB} and with NNs for X-ray spectroscopy~\cite{Rankine2020JPCA,Rankine2021JPCA} or for excitation spectra.~\cite{Ghosh2019AS} In the latter case, NNs could describe spectral intensities with deviations of 0.03 arb.u.. The same authors also fitted orbital energies of the QM9 data set comprising 134k organic molecules with a mean average error of 0.186~eV.~\cite{Ghosh2019AS} Alternatively, a diabatic\cite{Shu2020JCTC} or latent Hamiltonian matrix\cite{Westermayr2021physically} can be learned and used to obtain orbital energies or adiabatic energies as eigenvalues of the matrix, respectively. The latter approach was shown to improve the accuracy of orbital energy predictions by a factor of 2 compared to direct learning.\cite{Westermayr2021physically}}

{ML parametrization of excited states is especially challenging when multi-reference methods are required, because states can switch their character along certain reaction paths, which leads to jumps in the PESs. While this can also be the case for ground-state PESs, this problem is more pronounced for higher-lying excited states in regions where the density of states is high, leading to significant higher noise in excited-state PESs and consequently, more difficult learning.~\cite{Westermayr2019CS}} 

%TASK 3 improve accuracy with Delta-ML
%%%%%%%%%%%%%%%%%
{While ML parametrization of electronic structure data is well established, it is intrinsically limited in its application range} by the unfavorable scaling associated with bottleneck (1), \textit{i.e.}{,} many highly accurate electronic structure methods are too computationally costly to generate sufficiently large training datasets that enable reliable parametrization. Sometimes, \textbf{better accuracy can be achieved with $\Delta$-ML approaches}. This approach is based on the assumption that the difference in energy between two electronic structure methods - a low-level one and a high-level one -  is easier to represent than either one of the two methods.~\cite{ramakrishnan2015big} 
An alternative to the $\Delta$-learning approach is \textbf{transfer learning},~\cite{Pan2010IEEE} where a model is trained on data from a low level of theory and retrained with less data points of a more accurate method. A rule for determination of the number of data points needed in consecutive $\Delta$-learning approaches that takes computational cost and prediction accuracy into account is proposed by~\citet{Dral2020JCP}. Many studies use about 10\% of the original training data for \comment{$\Delta$-learning~\cite{Boselt2021JCTC,Ramakrishnan2015,Westermayr2021physically,Nandi2020JCP} and transfer learning.~\cite{Smith2019NC,ward2019machine,Kaeser2020JPCA,Kaeser2021arXiv,Qu2021breaking}}
In both cases, the ML model ideally yields an accuracy that is comparable to the higher-level theory. The prediction of energies with CC accuracy for the QM data sets was shown by \citet{Smith2017} using transfer learning and mostly range-separated semi-local DFT data {(5 million DFT data points compared to 500,000 CC data points)}. Very recently, \citet{Bogojeski2020NC} have demonstrated that {with $\Delta$-ML a model with CC accuracy was  generated} by using mostly semi-local DFT reference data and only a few data points calculated with CC theory. {For instance, MD of resorcinol (C$_6$H$_4$(OH)$_2$) could be achieved with 1004 data points at DFT and CC accuracy. While the DFT ML model had mean absolute errors of 2-3~kcal/mol compared to CC, the $\Delta$-ML model could achieve already 1~kcal/mol accuracy with respect to CC with as few as 25 data points.~\cite{Bogojeski2020NC}}

%TASK 4 get more physics included
Data efficiency can also be improved by designing NN architectures that implicitly satisfy symmetry constraints ({\textit{i.e.}}{,} rotational equivariance and permutational invariance) and, as a consequence, require much fewer data points to achieve a given accuracy.~\cite{batzner2021se3equivariant,schutt2021equivariant}
This is only one of many possible strategies to \textbf{include more physical information into ML model architectures}. Including the mathematical structures and the physical boundary conditions relevant to electronic structure methods into deep learning models leads to a further boost of data efficiency and model transferability. This has \comment{recently} been shown \comment{with reproducing kernels optimized for long-range intermolecular forces~\cite{Marko2006MP} and with} an ML-based parametrization of Density Functional Tight-binding (DFTB). \comment{The latter model provided} error reductions of up to 67\% for test molecules containing 8 heavy elements compared to existing DFTB parametrizations.~\cite{Li2018} Similarly, the MOB-ML approach uses localized 2-electron interaction integrals from Hartree-Fock calculations as input to construct a highly accurate and transferable GPR model. This is applied to the prediction of CCSD correlation energies for a diverse range of molecular systems.~\cite{Welborn2018,Cheng2019,husch2020improved} {The MOB-ML approach for instance reaches chemical accuracy by using three times fewer training data points for organic molecules with up to 7 heavy atoms compared to $\Delta$-ML approaches. Transferability was tested with molecules with up to 13 heavy atoms and MOB-ML could achieve chemical accuracy with 36 times fewer data points compared with $\Delta$-ML.\cite{Cheng2019}} 

Alternatively, rather than circumventing the solution of iterative equations of correlated wavefunction methods, ML models may also be used to facilitate faster convergence. {On average about 40\% reduction of the number of iterations for different basis sets could be achieved by \citet{Townsend2019JPCL}. They trained} an ML model to facilitate the convergence of CC methods based on lower-level theory electronic structure data. Besides ML models being powerful to accelerate the computation of target properties, they can also be used to predict correlated total energies of molecules based on Hartree-Fock or DFT results. Examples are NeuralXC,~\cite{Dick2020NC} DeepHC,~\cite{Chen2020_JPCA} and OrbNet~\cite{Quiao2020JCP} {which provide} NN representations based on atomic orbital features.

%ML to solve the many body problem
%Beyond ML representations of energy landscapes and properties as a function of chemical space, 
\textbf{ML becomes increasingly important as an integrated element of solving quantum many-body problems}. First attempts to solve non-homogeneous ordinary and partial differential equations using ML algorithms~\cite{Lagaris1997CPC,Lagaris1998CoRR,Sugawara2001,Manzhos2009} already date back to more than 20 years ago for model systems and have recently been applied to solve the quantum many-body problem for small organic molecular systems.~\cite{Carleo2017,Saito2017JPCJ,Nomura2017PRB,Han2018JCP,Pfau2020PRR,Hermann2020NC,Choo2018PRL,Zheng2019PRL}
These efforts have recently been summarized in a comprehensive review~\cite{Manzhos2020_review} and perspective.~\cite{Manzhos2020} While they are conceptually exciting and potentially transformative in solving the many body problem, their integration into existing, widely accessible electronic structure software may not be fully practicable yet as existing models are limited to small system sizes and {not yet} transferable.

%TASK ML to parametrize Hamiltonians in given basis representations
Rather than using ML methods to learn a representation of quantum states, they can also be used to parametrize electronic structure in an already known representation that is compatible with well-established electronic structure packages. Such \textbf{ML models are on their way to becoming an integrated element of electronic structure codes}. The resulting surrogate models, thereby, provide not only predictions of total energies and their derivatives, but further enable the derivation of many additional properties. 
%For example, FermiNet~\cite{Pfau2020PRR} %an antisymmetric NN model with accuracy of variational quantum Monte Carlo, or PauliNet~\cite{Hermann2020NC} %an ML model where the electronic structure is encoded in the nodes of the NN.
%are deep learning representations of the many body problem.
One such example is the SchNOrb model (SchNet for Orbitals),~\cite{schutt2019unifying} which is based on the deep tensor NN SchNet.~\cite{Schutt2018,Schuett2019JCTC} SchNOrb predicts Hamiltonians and overlap matrices in local atomic orbital representation compatible with most quantum chemistry software packages. Thus, it can be trained with data from quantum chemistry codes and its prediction can directly enter further quantum chemical calculations, {\textit{e.g.}}{,} as an initial guess of the wave functions in self-consistent field calculations or to perform perturbation theory calculations of correlation energies. {Self-consistent field iterations could be reduced by an average of 77\% when using the SchNOrb wave function as an initial guess.} Beyond that, it has been shown that the model can represent interaction integrals in localized effective minimal basis representations{, which benefits the prediction accuracy for larger systems}.~\cite{Gastegger2020JCP}

\begin{figure}[h]
    \centering
    \includegraphics[width=1.00\linewidth]{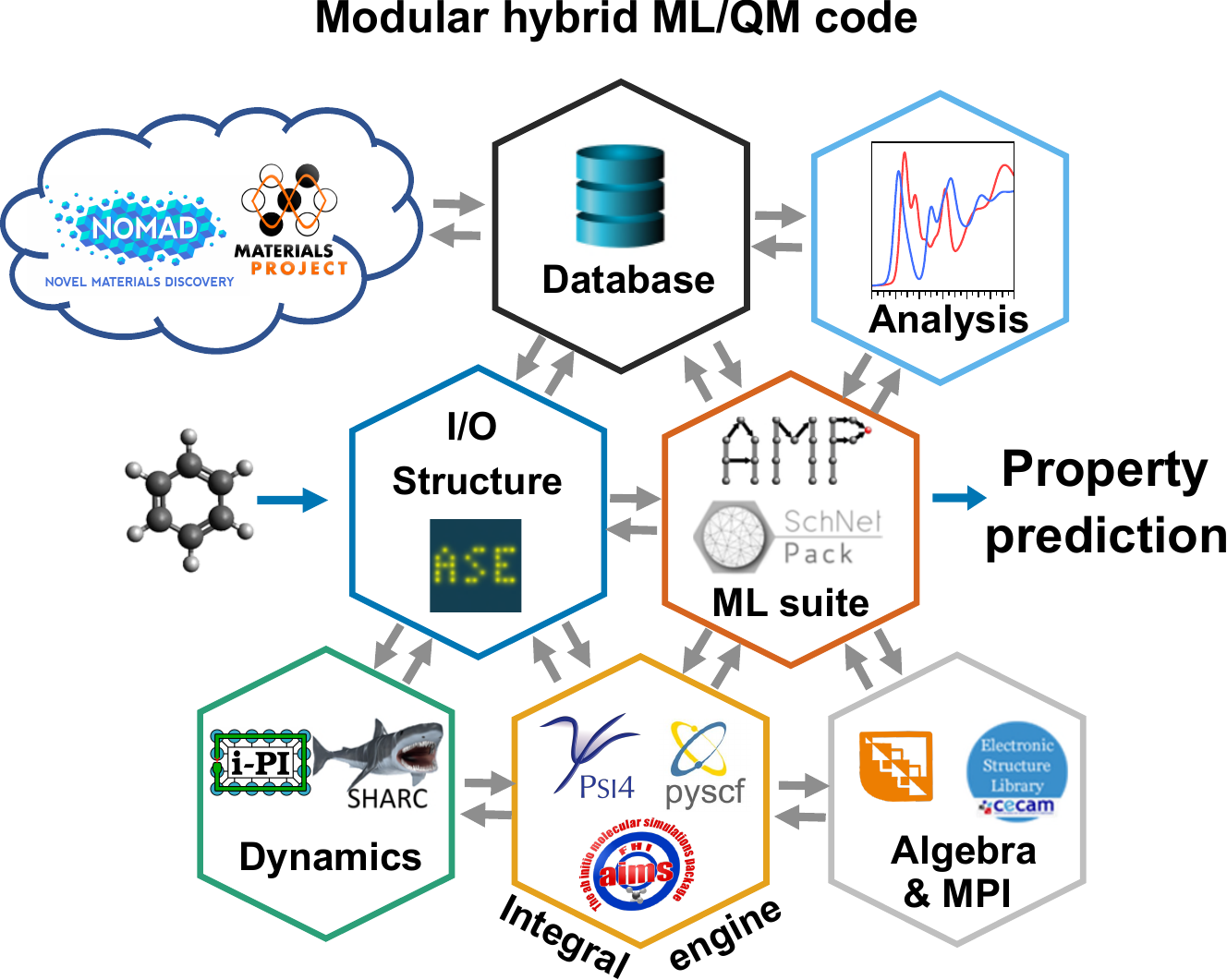}
    \caption{\label{fig:scf}Electronic structure software is increasingly becoming more modular. By moving away from monolithic (all-in-one) code models to a modular design, atomistic ML toolkits and data repositories, together with other standardized libraries, can be more {tightly} integrated into electronic structure workflows.}
\end{figure}

%TASK
Alternatively, an ML model may predict \textbf{the electron density or a density functional}.~\cite{Mahmoud2020PRB,Li2020PRL,Dick2020NC,Manzhos2020_review} A recent example of a deep learning framework to predict the electronic density or properties related to the density of a reference DFT method is DeepDFT.~\cite{Jorgensen2020arxiv_deepdft} 
{A symmetry-adapted method that considers geometrical covariance was proposed by \citet{Fabrizio2019CS} and \citet{Grisafi2019ACSCS} to learn the charge density of different organic molecules via Gaussian Process Regression (GPR) models.~\cite{Fabrizio2019CS,Grisafi2019ACSCS} This model is physically inspired and learns the charge density via a sum of atom-centered basis functions with the coefficients of these functions being predicted by the ML model. The authors achieve linear scaling with respect to the number of atoms and allow for size-extensive transferability. The latter was showcased by training the density of butadiene and butane and predicting the density of octa-tetraene and octane.\cite{Grisafi2019ACSCS} ~\citet{Fabrizio2020JCP} have further shown on the example of organic molecules that ML can be used to predict the on-top pair density in combination with a newly developed basis set. The on-top pair density can be used to assess electron correlation effects of a target compound, which most often cannot be described accurately using DFT. However, its evaluation requires post-HF or multi-reference calculations, which could be avoided due to the use of ML.}

A \textbf{universal density functional provided by an ML model} could potentially eliminate the need for exhaustively comparing different types of functionals for a given chemical problem. So far, ML has been used to generate new DFT functionals or to adjust the energy functional, bypassing the need to solve the iterative Kohn-Sham equations and accelerating simulations for the ground state~\cite{snyder2012finding,Brockherde2017, Babaei2020PRB,Schmidt2019JPCL,Nelson2019PRB,Lei2019PRM,Cheng2019,Dick2020NC} and excited states~\cite{Suzuki2020PRA} significantly. These models further promise better transferability for different types of molecular systems. Orbital-free DFT is another effort that allows for more reliable DFT calculations, but it requires the kinetic energy density functional.~\cite{Ligneres2005} However, various approaches have been put forward to parametrize the kinetic energy density functional with different kernel-based and deep learning methods.~\cite{Wang1999PRB,Golub2019PCCP,Seino2019CPL,Meyer2020JCTC}
\citet{Li2020PRL} recently presented an approach that integrates the iterative self-consistent field algorithm into an ML model to construct a learned  representation of the exchange-correlation potential for 1D model systems of H$_2$ and H$_4$.

%%semi empirical stuff, tight-binding etc.
The concept of ML-based Hamiltonian and density-functional surrogate models directly leads to the construction of \textbf{approximate electronic structure models based on ML}. Recently reported approaches include an ML-based H\"uckel model,~\cite{Zubatyuk2019} parametrized Frenkel~\cite{Farahvash2020JCP,Haese2016CS,zhang2020towards,Li2018,Kraemer2020JCTC} and Tight-binding (TB) Hamiltonians{~\cite{Wang2021}} as well as semi-empirical methods with ML-tuned parameters.~\cite{Dral2015,Chou2016JCTC}. 
Beyond that, several groups have proposed to combine established DFTB Slater-Koster parametrizations with kernel ridge regression or NN representations of the repulsive energy contributions to improve the accuracy and transferability of DFTB.~\cite{Stoer2020JPCL,Panosetti2020JCTC} {On the example of the QM7-X data set~\cite{Hoja2021SD}, a mean absolute error of 0.5 kcal/mol could be achieved on the atomization energies of the DFTB-ML model compared to hybrid DFT reference values.~\cite{Stoer2020JPCL}}%Electronic properties have further been obtained from self-consistent field with a NN layer that represents the tight-binding Hamiltonian with substantial error reduction for hydrocarbons.~\cite{Li2018}

{ \textit{Future directions:}
We expect a vivid development regarding the {tight} integration of ML within electronic structure software - an approach that some package developers already pursue ({\textit{e.g.}}{,} in the case of  \textit{entos}~\cite{Manby2019} and DFTB+~\cite{Hourahine2020}). Already in recent years, electronic structure software has started to move away from monolithic (all-in-one) software to more modular designs with interfaces to general-purpose standalone libraries\cite{CEN_Modular} (see Fig. \ref{fig:scf}). These developments will be helpful in the future to achieve integrated ML/QM solutions in computational workflows. As can be seen in Fig. \ref{fig:scf}, existing atomistic ML packages such as AMP,~\cite{Khorshidi2016CPC} sGDML~\cite{Chmiela2019CPC} or SchNetPack~\cite{Schutt2017_double,Schutt2018} could be interfaced with electronic structure packages that heavily expose internal routines ({\textit{e.g.}}{,} FHI-aims,\cite{Blum2009} PSI4,~\cite{Smith2020JCP} or PySCF~\cite{PYSCF}) and be used alongside dynamics packages such as i-Pi~\cite{Kapil2019CPC} and SHARC,~\cite{Mai2018WCMS,Richter2011JCTC} as well as algebra and electronic structure libraries such as ELSI~\cite{ELSI} and ESL.~\cite{ESL} The  structure generation, workflow and parser tool Atomic Simulation Environment (ASE)~\cite{Larsen2017IOPP} is for example already interfaced with the above examples of AMP and SchNetPack. This could also involve a closer integration with existing data repositories such as NOMAD,~\cite{Draxl2018,Draxl_2019} the Materials Project,\cite{Draxl2018,Draxl_2019} {the MolSSI QC Archive~\cite{Molssi2021}} or the \hyperlink{http://quantum-machine.org/datasets/}{Quantum Machine repository}.\cite{QMR} Universal data communication standards between quantum chemistry and ML will play an important role in the future. Efficient and scalable multi-language interoperability would further be needed to pursue the goal of tight integration of ML in electronic structure theory. In the future, we believe that ML will be part of many electronic structure codes to enable highly accurate electronic structure predictions at generally low computational costs. In this regard, data-efficient ML models are highly beneficial. Many recent works have shown that incorporation of symmetries and physical information into ML representations improves data efficiency, \textit{e.g.}, via the use of features derived from efficient low-level methods such as Hartree-Fock or MP2 theory to predict observables at high level of theories.~\cite{Welborn2018} Existing electronic structure software may further benefit from latent ML representations to mitigate existing bottlenecks in integral evaluations or to efficiently represent scalar and vector field quantities.}

\section{\label{sec:structure}ML will improve our ability to explore molecular structure and materials composition}

\begin{figure*}
\includegraphics[width=0.9\textwidth]{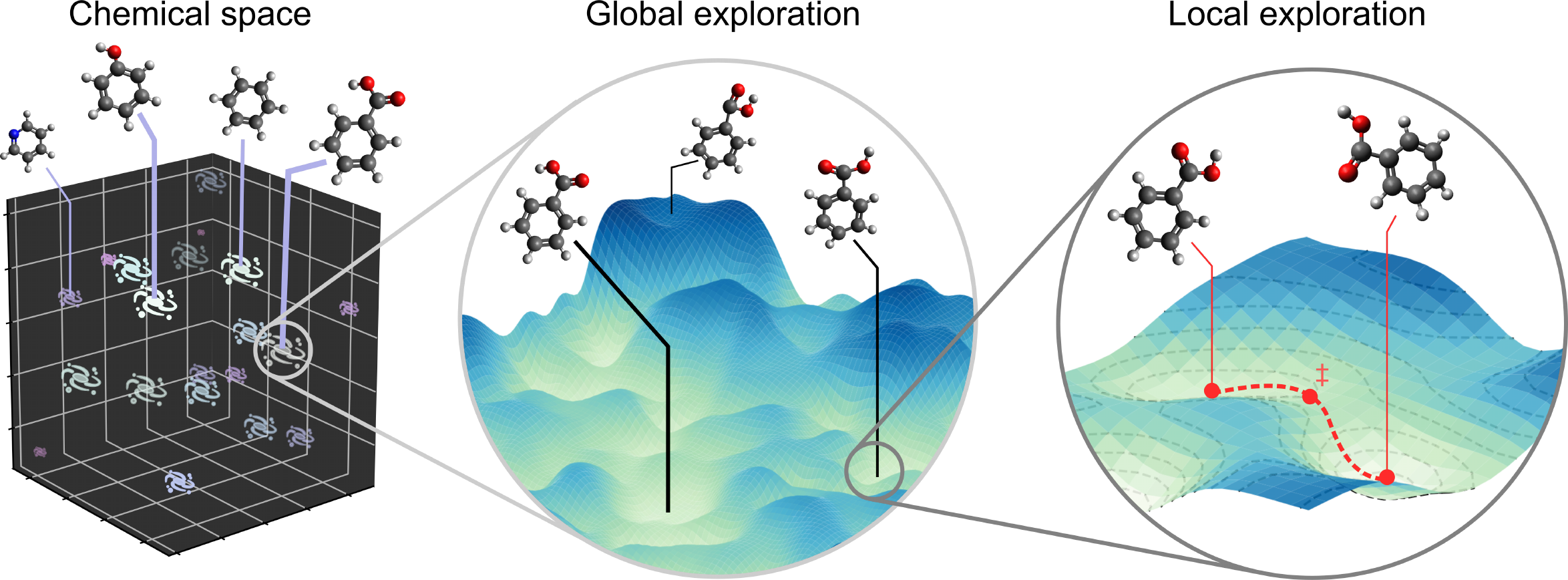}
\caption{\label{fig:pes_exploration}Exploration methods can target different scales of molecular and material space. 
At the highest level, chemical space, both chemical composition and structure are varied.
Global exploration targets a single {PES} with constant chemical composition and explores different structural conformations and their relative stability.
At the lowest level, local details of the {PES} such as reaction pathways and transition states are investigated.}
\end{figure*}

%general intro to chemical structure exploration
A key objective of computational chemistry and materials science is the prediction of new stable structures and viable reaction pathways to synthesize them. Beyond the significance to the discovery of new drugs and materials, finding stable equilibrium geometries and accessible transition states is a crucial element of computational molecular and materials discovery that typically involves tailored workflows.~\cite{Oganov2019} As shown in Fig. \ref{fig:pes_exploration}, optimization problems in atomistic simulation span different scales from searching stable molecules across chemical space to charting the global energy landscape spanned by the chemical coordinates of a given molecule down to local structure relaxation and transition state search. Even without considering the computational cost of electronic structure calculations, high-dimensional structure search is uniquely challenging and can be greatly facilitated by ML methods.

Efficient chemical exploration methods need to be able to identify CVs in high-dimensional spaces that are associated with relevant reaction events that occur at vastly different time scales ranging from the femtosecond regime (electron transfer and vibrational motion) to multiple nanoseconds (configurational dynamics of biomolecules) \cite{fleming1990chemical}. It is therefore not surprising that the use of a variety of methods that fall under the umbrella of ML, has led to a significant boost in the capability to explore chemical structure space. 

%local geometry optimization
Even a task that is nominally as simple as \textbf{finding the nearest equilibrium structure}, \textit{i.e.}{,} the local minimum of the potential energy landscape, can benefit from ML approaches.
The most common geometry optimization algorithms are based on quasi-Newton methods that determine trial steps based on an approximate Hessian. Finding optimal initial guesses and preconditioners for the Hessian is key to minimizing the number of geometry optimizations that are required. Recently, several more sophisticated preconditioning schemes have been proposed based on GPR that{, compared to established quasi-Newton algorithms,} significantly reduce the required number of geometry optimization steps for molecules and transition metal complexes~\cite{Denzel2018, Meyer2020,Raggi2020}, for correlated quantum chemistry methods that require numerical differentiation~\cite{Schmitz2018}, and for bulk materials and molecules adsorbed at surfaces.~\cite{PhysRevB.100.104103, GarijodelRio2020}
Furthermore, unsupervised ML can be used to automatically identify if geometry optimization has failed or led to an irrelevant outcome as recently shown for transition metal complexes.~\cite{Duan2020JPCL}

%reaction path sampling, transition state search, finding reaction coordinates
\textbf{ML methods have also recently been used to accelerate the search of first-order saddle points or transition states}. \citeauthor{Denzel2018} have used GPR {to speed-up} gradient-based transition state search starting from an equilibrium structure (one-ended search) {by a factor of 2 compared to conventional methods}.~\cite{Denzel2018,Denzel2020} Simultaneously, several approaches have been proposed to incorporate aspects of ML into double-ended transition state search based on the Nudged Elastic Band (NEB) method.~\cite{Peterson2016,Koistinen2017, GarridoTorres2019} \citet{GarridoTorres2019} have proposed a surrogate GPR model to accelerate a NEB method, leading to a factor of 5 to 25 fewer energy and force evaluations when compared to the conventional NEB method.

%global geometry optimization
One of the most challenging tasks, namely \textbf{identifying the global minimum of a potential energy landscape associated with the most stable structure}, can be significantly facilitated by the use of ML. Established methods to perform global optimization are often evolutionary algorithms or stochastic methods. {Examples for the former are} genetic algorithms~\cite{Curtis2018} and for the latter random structure search~\cite{Pickard2011} or basin hopping.~\cite{Doye_1997_BH,Panosetti2015} A prominent example of a global optimisation problem on a complex high-dimensional energy landscape is protein folding. Here, the alphaFold~\cite{alphafold} and alphaFold2~\cite{alphafold2} deep NN models were recently able to show what can be achieved when ML and structure optimisation methods are combined. In alphaFold, the ML model predicts residue distances and torsional angle distributions. On the basis of this, a coarse-grained potential is constructed to perform a sequence of random structure search and optimization cycles.
Hammer and coworkers have proposed a global structure prediction algorithm, called ASLA, based on image recognition and reinforcement learning.~\cite{Jorgensen2019,PhysRevB.102.075427} The use of image recognition to identify structural characteristics removes the need for encoding strings such as SMILES or descriptors of the atomic environment. The approach is applicable to molecules as well as materials and has been showcased on graphene formation, and oxide surface reconstructions.~\cite{Meldgaard2020} {In the case of graphene, the method is able to generate graphene as the most stable two-dimensional phase starting from initially random atom placement}.
Bayesian optimisation has become a common tool to achieve efficient structure prediction for crystals,~\cite{Yamashita2018,Deringer2018_faraday} surface reconstructions,~\cite{Bisbo2020} and hybrid organic/inorganic interfaces to name just a few examples.~\cite{Todorovic2019, Hormann2019} They often outperform evolutionary algorithms in terms of efficiency. 

%%%%CHEMICAL SPACE EXPLORATION
As shown in Figure \ref{fig:pes_exploration}, one level above the search for stable structures in energy landscapes lies the search for possible stable molecular compositions in chemical space. Generative ML models have recently shown great utility to predict molecules with tailored properties~\cite{Sanchez-Lengeling2018,schwalbe2020generative}, for example using SMILES  representation~\cite{gomez2018automatic} or  molecular graphs~\cite{liu2018constrained}.
While these are supervised approaches that require reference data for training, several related approaches have been proposed that use reinforcement learning.~\cite{Putin2018, Popova2018} These models can further be constrained to only predict SMILES strings that are chemically valid.~\cite{kusner2017grammar,Zhou2019}
Well beyond providing stability ranking, this approach can be used to generate molecules with arbitrary target properties to be used in drug and materials discovery. 
Unfortunately, {molecular} graph-based generative models are limited in their applicability, since they can not distinguish between different conformations that lead to the same graph.
However, for applications such as protein folding, optimizing reaction environments or finding reaction paths, it is paramount to have full access to conformation space.
\citet{mansimov2019molecular} proposed a generative model to sample 3d conformations from SMILES. This approach suffers from the same limitations as the graph representation it is built upon when properties are directly related to the 3d structure.
There have been several recent efforts to directly generate 3d molecular structures:
\citet{kohler2019equivariant} proposed equivariant normalizing flows, which are able to estimate a probability density over many-particle systems. This has been applied to finding meta-stable states of large Lennard-Jones systems.
\citet{gebauer_symmetry-adapted_2019} introduced G-SchNet that places atoms successively, incorporating rotational and translational symmetries. The model can be fine-tuned to generate molecules with properties in a specified target range.
%In a similar manner, \citet{simm_symmetry-aware_2020} have employed reinforcement learning to find stable molecules.

%%%TASK ALCHEMIC design
{\textit{Future directions}: 
With ML methods affecting every aspect of our ability to explore molecular configurations and compositions, their routine application to facilitate continuous exploration across composition space is not far, which would allow for the variation of the number and type of atoms in the system via \textbf{ML-enabled alchemical optimization}. So-called alchemical potentials have long been applied to rational drug design~\cite{von2005variational,von2007alchemical} and changing of reaction barriers.~\cite{sheppard2010alchemical} ML methods, such as NNs, have shown to be capable of modeling alchemical potentials~\cite{Faber2018,de2016comparing} as well as to produce smooth paths through alchemical space.~\cite{Schutt2016} We expect a lot of activity in this area in the future with ML methods enabling the continuous variation of elemental composition in materials to optimize their properties.
}

%%%%%%%%%%%%%%%%%%%%%%%%%%%%%%%%%%%%%%%%%%%
\section{\label{sec:QD} ML enables classical and quantum dynamics for systems of unprecedented scale and complexity}

%%%CLASSICAL DYNAMICS TASKS

%%%INTRO PARAGRAPH
The dynamical motion of atoms is a central target of a large part of computational research. In molecular simulation, we study the time evolution of electrons and atoms to predict static and dynamic equilibrium properties of molecules and materials at realistic temperature and pressure conditions, but also to understand nonequilibrium dynamics and rare events that govern chemical reactions. Dynamics methods range from  classical {MD}, via mixed quantum-classical dynamics (MQCD) methods (incorporating electronic quantum effects) to quantum dynamics in full quantum or semi-classical formalisms. In all cases, equations of motion need to be integrated over time, which involves numerous evaluations of forces and other properties that govern the dynamics. ML methods can address bottlenecks in such simulations on various levels: {Their most prevalent use is to speed up energy, force, and property evaluations in each time step by providing ML-based force fields and interatomic potentials. Other ML approaches directly target MD by supporting coarse-graining and the use of larger time steps, or by replacing MD completely with a direct prediction of dynamical properties, expectation values, and correlation functions.}

%%TASK 1 facilitating equilibrium dynamics and the simulation of static/dynamic properties 
The most obvious way in which ML can facilitate MD simulations is the \textbf{use of ML-based interatomic potentials instead of on-the-fly \emph{ab-initio} MD}. {Many early applications of ML in molecular simulation were mostly focused on ML parametrization of electronic structure data for the benefit of MD simulation.} ML-based interatomic potentials that replace electronic structure evaluation during dynamics {are} by now commonly established, see{,} {\textit{e.g.}}{,} Refs.~\citenum{Behler2016},~\citenum{Unke2020arXiv}, and \citenum{Botu2017JPCC}, and {have} since enabled simulations of unprecedented complexity and scale. For example, a recent breakthrough by \citet{Deringer2021N} showed that Gaussian Approximation Potentials\cite{Bartok2010,de2016comparing} could be used to predict phase transitions and electronic properties of systems containing more than 100,000 atoms. \citet{Jiang2020JPCL} have recently reviewed the transformative role that ML-based high-fidelity PESs play in gas-surface dynamics simulations. 

\comment{In principle, approaches can be distinguished between those that sample molecule deformations around an equilibrium geometry, e.g. for optimizations,~\cite{Meyer2020JCP} or those that consider "reactive" potential energy surfaces.~\cite{Unke2020MLST,Koner2019JCP,Danielsson2008JCTC,Bowman2011PCCP,Jiang2020JPCL,Meuwly2021arXiv} An alternative approach is to directly predict targeted simulation properties such as reaction yields.~\cite{Haese2019CS,Houston2019JPCL}}
A key factor in building ML force fields {for MD simulations} is the efficient and comprehensive sampling of relevant data points. Active learning schemes have been proposed\comment{~\cite{Li2015PRL,Botu2015PRB,Behler2015IJQC,gastegger2017machine,Akimov2018JPCL,Westermayr2019CS}} to efficiently sample the relevant configuration space {that a molecule visits during an} MD simulation. These schemes are based on an uncertainty measure during ML dynamics, which can be used to detect unexplored or undersampled conformational regions. \comment{The uncertainty measure could be for instance the deviation of two NNs or the statistical uncertainty estimate of the inferences made with{,} {\textit{e.g.}}{,} GPR. One way to measure the accuracy and interpolative regime of ML models is to use the previously mentioned adaptive sampling techniques also during the production runs. This allows to detect holes in the potential energy surfaces \emph{on-the-fly}.~\cite{Westermayr2019CS}}

By using gradient-domain ML models that are trained on gradients rather than energies, energy-conserving ML force fields can be obtained with high accuracy and little amount of training data required.~\cite{chmiela2017,Chmiela2018NC,Chmiela2019CPC} {$\Delta$-ML models, in the context of MD simulations,  have also proven to be very powerful in providing a data-efficient representation of CC accuracy from DFT data~\cite{Bogojeski2020NC} or DFT accuracy from mostly DFTB data in the context of QM/MM simulations~\cite{Boselt2021JCTC}, to name two recent examples.} 
%TASK 2 
%ML helps with studying the dynamics of rare events for chemical reaction dynamics, folding etc.
{Beyond the use of ML to facilitate accurate force evaluations in MD,} ML methods {have been used to} enable the simulation of rare events that occur on time scales inaccessible to conventional MD. A perspective review that recently arose from a CECAM conference on "Coarse-graining with ML in molecular dynamics" provides a comprehensive overview of ML for free energy sampling, coarse-graining, and long-time MD~\cite{Gkeka2020}. 

%dimensionality reduction / collective variables
\textbf{ML methods help to identify CVs, which characterize long-time dynamics} of molecular systems. This is important to identify long-lived attractor states in phase spaces and to find strategies to efficiently explore dynamics in complex hierarchical energy landscapes, {\textit{e.g.}}{,} for \comment{ the isomerization of alanine dipeptide~\cite{Ma2005JPCB} or for} protein folding.~\cite{Noe2020a}
ML methods in this domain based on principal component analysis~\cite{pearson1901lines} date back to over 20 years ago.~\cite{Balsera1996} \comment{More recent approaches include} kernel principal component analysis~\cite{scholkopf1998nonlinear,Zhang2008,Lange2008}, diffusion maps,~\cite{Coifman2006,Preto2014,Zheng2013}, \comment{the Sketch map method,~\cite{Tribello2012PNAS,Ceriotti2011PNAS}} Markov state models~\cite{Mardt2018,Noe2019} and various types of autoencoders.~\cite{Chen2018,Ribeiro2018}

%TASK 3 ML helps with bridging scales and developing coarse-grained models
Several ML models have been developed that aim to achieve \textbf{bottom-up coarse-graining} by representing the potential of mean force or free energy surface as a function of coarse grained variables. This has been done for instance using NNs to infer conformational free energies for oligomers~\cite{Lemke2017} or to construct a coarse-grained liquid water potential~\cite{Zhang2018} or using a Gaussian approximation-based coarse-grained potential for alanine dipeptide~\cite{wang2020} and molecular liquids.~\cite{John2017}
%%%%%%%%%%%%%%%%%%%%%%%%%%%%%%%%%%%%%%%%%

%PARAGRAPH CLASSICAL DYNAMICS TO MQC Dynamics
MQCD, {\textit{i.e.}}{,} classical dynamics of nuclei coupled to the time-dependent quantum mechanical evolution of electrons, are commonly used to simulate light-induced nonadiabatic dynamics of molecules,~\cite{Barbatti2011,gonzalez2020quantum,Mai2020ACIE} as well as coupled electron-nuclear dynamics in extended systems.~\cite{Smith2020_JPCM} While on-the-fly MQCD simulations have become feasible in the last decade, the accessible time scale and the number of non-equilibrium trajectories that can {realistically} be simulated on-the-fly is too limited to enable comprehensive statistical analysis and ensemble averaging. \textbf{ML shows great promise in nonadiabatic excited-state simulations}\cite{Westermayr2020CR,Westermayr2020MLST_Perspective} as documented by recent works using NNs to construct excited-state energy landscapes to perform fewest-switches surface hopping MD at longer time scales or with more comprehensive ensemble averaging {than would otherwise be possible with on-the-fly dynamics}.~\cite{Westermayr2020JPCL,Westermayr2019CS,Li2020chemrxiv} Similar progress has been achieved in nonadiabatic dynamics at metal surfaces, where NNs have been used to construct excited-state landscapes~\cite{Carbogno2008PRL,Carbogno2010PRB} and continuous representations of the electronic friction tensor~\cite{Zhang2020_friction} used in {MD} with electronic friction simulations.~\cite{C8SC03955K,Box2020JACSAu} 

%QUANTUM DYNAMICS
Even \textbf{full quantum dynamics simulations} have recently seen an increasing uptake of ML methodology to push beyond longstanding limitations in the dimensionality of systems that can be simulated. The main bottleneck in quantum dynamics simulations is not the evaluation of the temporal evolution of the electrons, but the temporal evolution of the nuclear wavefunction, which involves computations that (formally) scale exponentially with the number of atoms in the system. Potential energy landscapes in quantum dynamics are typically represented in a diabatic basis rather than the adiabatic representation (directly outputted by electronic structure codes) in a process called  (quasi-)diabatization.~\cite{Yarkony2012CR,Koeppel2004} However, quasi-diabatization requires expert knowledge and is highly complex for more than two coupled electronic states. The construction of diabatic representations with deep NNs has recently shown {great} potential to simplify and automate this laborious task.~\cite{Shu2020JCTC,Jiang2016IRPC,Lenzen2017JCP,Williams2018JCP,Xie2018JCP,Williams2020JPCA}
Besides the PES generation itself, recent works use GPR to fit the diabatic PESs in reduced dimensions.~\cite{Richings2017CPL,Richings2018JCP,Richings2019JCTC,Richings2020JCP} One of the largest ML-enhanced quantum dynamics simulation was {recently} performed on a 14-dimensional energy landscape for a mycosporine-like amino acid~\cite{Richings2019FD}.

\comment{The computational efficiency of ML models is an important point to consider. MD simulations based on ML models are considerably more efficient than \emph{ab initio} MD, yet still relatively slow compared to empirical force fields. For example, 100 femtosecond MQCD MD of CH$_2$NH$_2^+$ on a single compute core take 24 seconds with ML potentials compared to 74,224 seconds with the reference method (MR-CISD/aug-cc-pVDZ).~\cite{Westermayr2020JPCL} The simulation of 100 femtosecond classical MD of the same molecule in the gas phase with an empirical force field takes 0.005 seconds with Amber.~\cite{salomon2013overview} The computational efficiency of ML models can become a bottleneck if long time scales or ensemble averages over many thousands of reaction events are required. Similar memory and CPU efficiency bottlenecks can arise during model training of kernel methods and deep neural networks if large training data sets and complex high dimensional models are involved. }

%last par
{\textit{Future directions}: 
ML-based interatomic potentials and continuous regression models already play an important role across almost all domains of MD simulations and we expect that the use of ML in MD will further increase in the coming years. \comment{As larger and more complex systems are targeted and longer time scales are needed, a future challenge that needs to be tackled is the computational efficiency of ML models, especially for MD simulations. The concept of sparsity in terms of ML methods and data representation can lead to better computational efficiency. Recently, explicit atomic high body order expansions in permutationally invariant polynomials (\textit{e.g.} aPIPs\cite{oordRegularisedAtomicBodyordered2020}, ACE\cite{drautzAtomicClusterExpansion2019}) have emerged as appealing alternative to kernel and deep learning methods as they accurately allow high-dimensional parametrization as a function of atomic coordinate spaces and can be trained by linear regression. As a result, both training and evaluation are highly efficient with evaluation times on the order of few milliseconds per atom.~\cite{allenAtomicPermutationallyInvariant2021a}} 
While most approaches focus on assisting MD by providing highly-accurate interatomic potentials and force fields, they have also shown great potential in predicting dynamical properties directly and skipping the MD simulation completely or in assessing the validity of different approximations in dynamical simulations. The latter has only recently been shown by \citet{Jasinski_2020} with a Bayesian model to estimate errors due to different approximations in quantum scattering simulations.
Going forward, complex dynamical simulation methods will become more accessible to non-expert users with the help of ML and will open avenues to tackle complex systems in solvent environments~\cite{Chen2019JPCL} or dynamics at hybrid organic-inorganic interfaces.~\cite{Box2020JACSAu}  It is evident that ML methods will play an important role in extending the range of applications for MQCD methods in the coming years. A recent work by \citet{Brieuch2020JCP} employing ML methods to achieve converged path-integral MD simulations of reactive molecules in superfluid helium under cryogenic conditions is an exemplary showcase of what the synergy of ML and quantum dynamics methods can achieve.
}

%%%%%%%%%%%%%%%%%%%%%%%%%%%%%%%%%%%%%%%%%%%%%%%
\section{ML helps to connect theory and experiment}~\label{sec:exp}

\begin{figure}
\includegraphics[width=0.9\columnwidth]{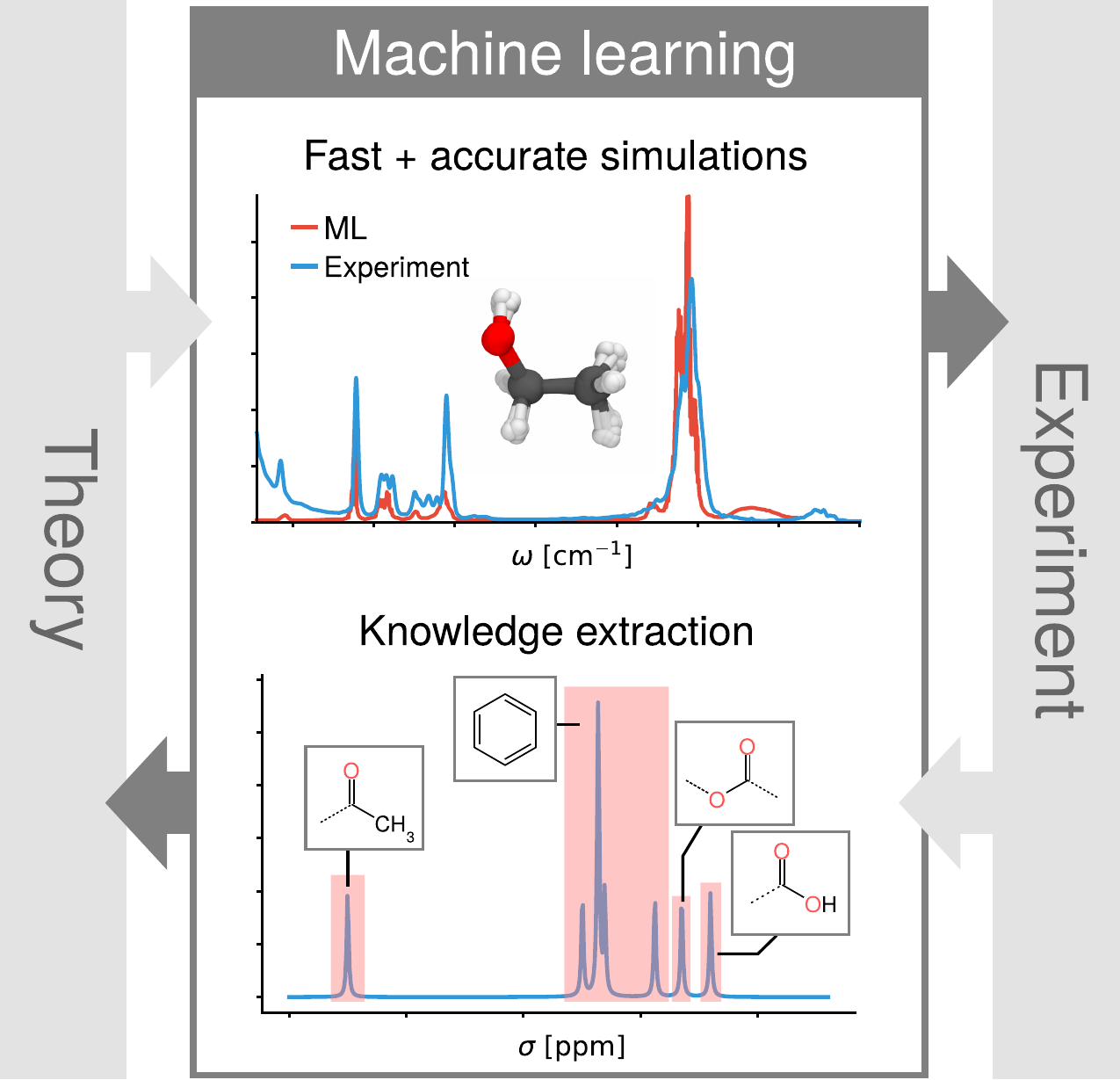}
\label{fig:experiment}
\caption{Depiction of how ML methods can act as a bridge between theory and experiment. ML models trained on theory predict spectra with realistic lineshapes. At the same time, ML models can be used to infer structural information from experimental measurements.}
\end{figure}

% General task
The ultimate goal of computational molecular and materials simulation is to connect theory and experiment. This could mean supporting the explanation of experimental outcomes or finding new theoretical rules in observations, in both cases leading to a {better} understanding of the physical world and its laws. Forming this connection is a hard task. A plethora of different effects need to be considered in even the simplest atomistic systems, making it very difficult to faithfully reproduce experimental conditions in silico. On the other hand, experimental observations can be obscured by a variety of influences or by the sheer complexity of the measured signal. % Then the search for new insights becomes the figurative search for a needle in the haystack.
% ML context
%ML can provide an important contrib in that respect. 
As we have seen in the preceding sections, \textbf{ML approaches can increase the accuracy of predictions and the speed with which they can be obtained.}
This makes it possible to carry out computational studies which close the gap between theory and experiment by more efficiently incorporating experimental parameters such as finite temperature, measurement conditions, and solvent effects.
Moreover, ML techniques can also provide invaluable support in extracting information from experimental observations and uncovering trends that are not directly apparent to the  practitioner.

One field which has greatly profited from these developments is \textbf{computational spectroscopy}. The prediction of spectroscopic properties is a central aspect of computational modeling, as it provides results which can be directly compared against experiments. Examples of successful ML applications include the prediction of different vibrational spectra, combined with different response properties of the electric field. \citet{gastegger2017machine} have combined a latent charge dipole model with interatomic potentials in order to efficiently simulate infrared spectra (IR) of organic molecules in gas phase {without having to resort to electronic structure computations of the molecular dipole}. This approach has further been applied to model absorption spectra.~\cite{zhang2020towards,Westermayr2020JCP} \citet{raimbault2019using} introduced a kernel approach for predicting the Raman spectra of organic crystals based on molecular polarizabilities. Using a NN based approach, \citet{sommers2020raman} have demonstrated that ML can also be used to simulate Raman spectra of extended systems such as liquid water{, which would be computationally unfeasible when done with DFT}.
In addition to vibrational spectra, ML models are also capable of modeling response properties, allowing the simulation of electronic excitations using{,} {\textit{e.g.}}{,} MQCD approaches (see Section~\ref{sec:QD}). For example, \citet{zhang2020towards} use NN models to obtain transition dipole moments, which in turn could be used to predict UV and visible light spectra.
ML approaches have further been used to predict nuclear magnetic resonance (NMR) spectra from molecular simulations. \citet{paruzzo2018chemical}, for example, have used the kernel model from Ref.~ \citenum{raimbault2019using} to predict the chemical shifts in molecular solids. 
Recently, Christensen \emph{et al.} have introduced an electric field dependent descriptor in the FCHL Kernel framework\cite{Christensen2019}. Based on this, they have derived molecular dipole moments as a general response to the electric field, which can be used to simulate IR spectra of small organic molecules.
\citet{gastegger2020machine} have applied a response theory approach in combination with a deep NN architecture which explicitly depends on electric and magnetic fields. They could show that, in this manner, a single ML model can predict IR, Raman and NMR spectra. Moreover, by introducing the field generated by a molecular environment they were able to model the effect of solvents on the resulting spectra.

% ML for extraction
{Beyond that, ML offers the possibility to \textbf{directly extract information from experimental observations} and relate them to fundamental chemical concepts.
% interpretation 
One example is the use of ML to interpret different types of spectroscopic measurements to determine structural or electronic properties of molecules and materials.} \citet{fine2020spectral} have recently presented a ML approach to extract data on functional groups from infrared and mass spectroscopy data, while \citet{kiyohara2018data} have successfully applied a ML scheme to obtain chemical, elemental, and geometric information from the X-ray spectra of materials. Another application where ML shows promise is the automated interpretation of nuclear magnetic resonance spectra with respect to atomic structure, which typically relies heavily on experience.~\cite{cobas2020nmr}

% data mining
However, \textbf{ML can also be used to leverage {information contained in large collections of scientific data}}. The majority of chemical knowledge is collected in the form of publications. ML approaches such as natural language processing and image recognition offer the possibility to directly distill functional relationships and chemical insights from the massive body of scientific literature. For instance, \citet{tshitoyan2019unsupervised} have used natural language processing to extract complex materials science concepts, such as structure property relationships, from a large collection of research literature. They could further demonstrate, that their model was able to generalize on the learned concepts and recommend materials for different functional applications. \citet{raccuglia2016machine} recently trained a ML model using information on failed experiments extracted from archived laboratory notebooks to predict the reaction success for the crystallization of templated vanadium selenites. Their model was able to learn general reaction conditions and even revealed new hypotheses regarding the conditions for successful product formation.

Finally, ML offers new ways in which theory can guide experiment.
Two fields where ML has played a transformative role are
\textbf{molecular/materials discovery and computational high-throughput screening}, with several reviews summarizing recent advances.~\cite{Schleder2019, Elton2019, Yang2019CR, Goldsmith2018,Toyao2020_ACS_Catalysis,McCullough2020}
The combination of high-throughput screening with accurate and efficient ML models has proven to be highly valuable, as it allows to substitute most of the required electronic structure calculations~\cite{melville2009machine}.
Examples of what is possible in this space include the objective-free exploration of light-absorbing molecules,~\cite{Terayama2020} drug design,~\cite{ekins2019exploiting} the computational search for highly active transition metal complexes that catalyse C-C cross coupling reactions,~\cite{Meyer2018} or the discovery of new perovskite materials~\cite{GomezPeralta2020} or polymers for organic photovoltaic applications.~\cite{jorgensen2018machine, StJohn2019}

Still, chemical space is estimated to cover more than $10^{60}$ molecules~\cite{Dobson2004N}, hence  exhaustive computational screening remains infeasible -- even with fast and accurate ML models.
In this context, \textbf{ML-enabled inverse design} offers a promising alternative by reversing the usual paradigm of obtaining properties from structure~\cite{Weymuth2014,zunger2018inverse}.
Instead, the aim is to create structures exhibiting a range of desired properties.
Since such ML models readily provide analytic gradients, an application to property-based structure optimization is straightforward.
First steps of applying ML in these areas have recently been achieved.
Examples include the optimization of the HOMO-LUMO gap as demonstrated by \citet{schutt2019unifying} and relaxation for crystal structure prediction as investigated by \citet{podryabinkin2019accelerating}.
While ML only provides gradient-based local optimization in these examples, it can be combined with genetic algorithms~\cite{podryabinkin2019accelerating} or global optimization methods such as simulated annealing or minima hopping~\cite{noh2020machine}.

\textit{Future directions}: 
While ML techniques and atomistic ML potentials in particular have contributed greatly to  closing the gap between theory and experiment, a range of open issues remains.
Problems that have only recently begun to be studied include how to extend ML simulations to efficiently reproduce different experimental conditions, such as solvents or electromagnetic fields.
Another frequently encountered issue concerns the data efficiency of ML models, as well as the availability of reliable reference data.
% Generative models and inverse design approaches, for example, to date primarily target computed properties as stand in for more complex experimental data that can either be hard to obtain or be subject to experimental uncertainty and measurement noise.
\comment{For example, most generative models and inverse design approaches to date primarily target simulated properties rather than experimentally measured ones.
While calculated quantities (\textit{e.g.} redox potentials, singlet-triplet gaps) can offer invaluable guidance for design endeavors, they ultimately represent approximations to the physical characteristics of a system, which can only be fully captured through experiments (\textit{e.g.} full-cell study for redox kinetics and electrochemical stability).
Successful design endeavors therefore often combine theoretical computations with experimental data or calibrate against them\cite{gomez2016design, lin2016redox}.}

%%%FINAL CHAPTER
\section{Outlook}

%from past to present
%what deep integration can achieve
We expect that ML methods will soon become {an integrated} part of electronic structure and molecular simulation software pushing the boundaries of existing techniques {toward} more computationally efficient simulations. ML methods may for example replace complex integral evaluations in the construction of Hamiltonians and secular equations or they can provide improved initial guesses to iteratively solve integro-differential equations. ML methods can further help to describe non-local effects in time and space and provide mechanisms for on-the-fly uncertainty quantification and accuracy improvements. The beneficial scaling properties of ML algorithms with respect to the size of atomistic systems will play an important role in extending the range of application of existing electronic structure and dynamics simulation methods. The application of ML to MQCD simulations will {make it possible to reach} currently unfeasible time and length scales {beyond few picoseconds and tens of atoms. This will in turn require the improvement of existing molecular simulation methods to capture long time dynamics.} As we explore systems of increasing size, we will be able to better study the boundary between quantum effects at the nanoscale as well as collective many-body effects and fluctuations at the meso- and macroscale.~\cite{Ambrosetti1171}

%we need new software
A necessary requirement is the establishment and the distribution of user-friendly and well-maintained \textbf{simulation software with {tight} integration of ML methodology} in chemistry and materials science. Software solutions will need to be modular to allow interfacing with well-established deep learning platforms such as TensorFlow or PyTorch. This should involve the establishment of common data standards to easily communicate atomistic simulation and electronic structure data between chemistry and ML packages. In many ways, this requirement is in line with recent trends of increased modularity of codes via general libraries such as ESL~\cite{ESL} and ELSI~\cite{ELSI}  (see Fig.~\ref{fig:scf}). A {recent} initiative {toward} an integration of ML is the ENTOS quantum chemistry package and ENTOS AI~\cite{Manby2019}. 

%we need data sharing
Another challenge ahead is related to \textbf{establishing a culture of openness and willingness to share data and ML models} as the availability of training data is a crucial aspect of driving advances in this field. {While data sharing is quite common in material science, it is not yet so common in computational molecular science.} Well defined materials data standards as put forward by the Fair Data Infrastructure project (\hyperlink{https://www.fair-di.eu/fairmat/}{FAIR-DI})\cite{Wilkinson2016SD} and \emph{ab-initio} data repositories such as for example the NOMAD repository\cite{Draxl2018,Draxl_2019}, the Materials Project\cite{Jain2013}{, and the MolSSI QCArchive~\cite{Molssi2021}} are needed {in all research areas}. The need for open access to vast amounts of data will need to be balanced against other needs, such as commercial interests that arise from industrial research or commercial software projects.

%we need community and funding
Sustainable integration of ML methods into widely-used software will require long-term community effort and might be less glamorous than exciting proof-of-principle applications of ML in chemistry and materials science. Research funding agencies, reviewers, and industrial stakeholders need to acknowledge this and ensure that sustained funding for such efforts is put in place.

%Outcomes: change of workflow, change of code design
If achieved, {an integration} of ML methodology into electronic structure and molecular simulation software, will induce lasting change in workflows and capabilities for computational molecular scientists. Furthermore, it will offer the opportunity to reconsider many of the underpinning design choices of electronic structure and molecular simulation software packages which, in many cases, historically arose from computational efficiency considerations. For example, Gaussian basis representations have been chosen decades ago in quantum chemistry due to the ease of evaluating multi-centre integrals. If ML methods can vastly facilitate the evaluation of multi-centre integrals, are Gaussian basis functions still the best choice of basis representation? 

%outcome: widen participation
{An i}ntegration of ML and molecular simulation will drastically widen participation in the field and uptake of our methods and problem solving approaches. If codes require dramatically fewer computing resources and offer the ability to directly predict experimentally accessible quantities, computational simulation will become more appealing as a complementary tool in synthetic and analytical labs. In many industrial applications, cost-benefit analysis requires that a clear correspondence exists between the cost of delivering predictions and the accuracy and precision that is required for an application. {The use of} ML methods {within such workflows} will hopefully also provide a drive {toward} establishing better measures of uncertainty in atomistic simulation.

%outcome: change of training
Finally, \textbf{the method portfolio and skill set of computational molecular scientists will need to adapt} as a consequence of the growing importance of ML methods in electronic structure theory and molecular simulation. In many cases, the presence of some aspects of ML "under the hood" of existing methods and workflows will not change how we apply these methods. For example, a DFT functional parametrized by a ML approach, can be applied as any existing functional (although its range of applicability might be very different). In other cases, the presence of ML methods will fundamentally change basic workflows as we have discussed across the sections of this perspective. In those instances, practitioners need a basic understanding of ML concepts and the different models that they are working with. This involves knowledge of the capabilities and limitations of most standard applications to avoid pitfalls. As such, ML methodology will have to become an integral part of education in computational chemistry and materials science.

\begin{acknowledgments}
This work was funded by the Austrian Science Fund (FWF) [J 4522-N] (J.W.), the Federal Ministry of Education and Research (BMBF) for the Berlin Center for Machine Learning / BIFOLD (01IS18037A) (K.T.S.),  and the UKRI Future Leaders Fellowship programme (MR/S016023/1) (R.J.M.).
M.G. works at the BASLEARN – TU Berlin/BASF Joint Lab for Machine Learning, co-financed by TU Berlin and BASF SE.
\end{acknowledgments}

\section*{Data Availability Statement}
Data sharing is not applicable to this article as no new data were created or analyzed in this
study.

%\bibliography{jcp_perspective}

\section*{References}
%aipnum4-2.bst 2019-01-14 (MD) hand-edited version of apsrev4-1.bst
\end{document}